\documentclass[aps,prl,twocolumn,showpacs,superscriptaddress]{revtex4-1}  % for review and submission
\usepackage{orcidlink}
\usepackage{graphicx}  % needed for figures
\usepackage{dcolumn}   % needed for some tables
\usepackage{bm}        % for math
\usepackage{amssymb}   % for math
\usepackage{lipsum}
\usepackage{xcolor}
\usepackage{multirow}
\usepackage[autostyle]{csquotes}
\usepackage{graphicx}
\usepackage{subfigure}
\usepackage[leftcaption]{sidecap}
\usepackage{soul}
\begin{document}
% The following information is for internal review, please remove them for submission
%\linenumbers
\widetext
%\leftline{Version 09 as of \today}
%\leftline{Primary authors: $^1$M. Kumar, $^2$V. Bhardwaj, $^1$K. Lalwani}
%\leftline{Institutes: $^1$MNIT Jaipur, $^2$ IISER Mohali}
%\leftline{Belle Note number :BN1622}
%\leftline{To be submitted to PRD(L)}
%\leftline{Comment to {bn1622\_ml@bpost.kek.jp}}

% the following line is for submission, including submission to the arXiv!!
%\hspace{5.2in} \mbox{Fermilab-Pub-04/xxx-E}
%%% Paper:    B+ to D_s(*)+ eta etc
%%% Journal:  Physical Review D Letter
%%% Contacts: M. Kumar (2016RPY9052@mnit.ac.in)
%%%           V. Bhardwaj (vishstar@gmail.com)
%%%           K. Lalwani (kavita.phy@mnit.ac.in)
%%% Non-responding authors or those who said NO are commented out.
%%% ====================================================================
%%% Click the RELOAD button on your web browser to see the updated file.
%%% ====================================================================
%%% Use \input{author-orcid} to insert this material into your latex file.
%%%%% Force institutions to appear in alphabetical order when typeset.
\noaffiliation
  \author{M.~Kumar\,\orcidlink{0000-0002-6627-9708}} % 2744
   \author{V.~Bhardwaj\,\orcidlink{0000-0001-8857-8621}} % 2228
   \author{K.~Lalwani\,\orcidlink{0000-0002-7294-396X}} % 2142
  \author{I.~Adachi\,\orcidlink{0000-0003-2287-0173}} % 2590
% \author{K.~Adamczyk\,\orcidlink{0000-0001-6208-0876}} % 2239
% \author{J.~K.~Ahn\,\orcidlink{0000-0002-5795-2243}} % 7423
  \author{H.~Aihara\,\orcidlink{0000-0002-1907-5964}} % 2223
% \author{S.~Al~Said\,\orcidlink{0000-0002-4895-3869}} % 6823
  \author{D.~M.~Asner\,\orcidlink{0000-0002-1586-5790}} % 4684
% \author{H.~Atmacan\,\orcidlink{0000-0003-2435-501X}} % 2538
% \author{V.~Aulchenko\,\orcidlink{0000-0002-5394-4406}} % 8183
  \author{T.~Aushev\,\orcidlink{0000-0002-6347-7055}} % 3747
% \author{R.~Ayad\,\orcidlink{0000-0003-3466-9290}} % 3766
% \author{T.~Aziz\,\orcidlink{-}} % 3523
  \author{V.~Babu\,\orcidlink{0000-0003-0419-6912}} % 5623
% \author{S.~Bahinipati\,\orcidlink{0000-0002-3744-5332}} % 2332
% \author{A.~M.~Bakich\,\orcidlink{0000-0001-8315-4854}} % 2115
% \author{Y.~Ban\,\orcidlink{-}} % 3503
% \author{E.~Barberio\,\orcidlink{-}} % -229
% \author{M.~Barrett\,\orcidlink{0000-0002-2095-603X}} % 2180
% \author{M.~Bauer\,\orcidlink{0000-0002-0953-7387}} % 9863
  \author{P.~Behera\,\orcidlink{0000-0002-1527-2266}} % 4204
  \author{K.~Belous\,\orcidlink{0000-0003-0014-2589}} % 2329
% ...
% 2387
% \author{V.~Zhulanov\,\orcidlink{0000-0002-0306-9199}} % 4983
\collaboration{The Belle Collaboration}

\title{Search for rare decays  $B^{+} \to D_{s}^{(\ast)+}\eta$, $D_{s}^{(\ast)+}\bar{K}^{0}$, $D^{+}\eta$, and $D^{+}K^{0}$}
%\collaboration{The Belle Collaboration}
%\date{\today}

\begin{abstract}
We present a study of rare decay modes $B^{+} \to D_{s}^{+}h^{0}$, $B^{+} \to D_{s}^{\ast+}h^{0}$, and $B^{+} \to D^{+}h^{0}$, where $h^{0}$ denotes the neutral meson $\eta$ or  $K^{0}$, using a data sample of $(772 \pm 10 ) \times 10^{6}$ $B\bar{B}$ events produced at the $\Upsilon(4S)$ resonance. The data were collected by the Belle detector operating at the asymmetric-energy KEKB collider. We find no evidence for these decays, so we set upper limits at the 90$\%$ confidence level on the branching fractions of $B^{+} \to D_{s}^{+}h^{0}$, $D_{s}^{\ast+}h^{0}$, and $D^{+}h^{0}$ decay modes. Along with these rare decay modes, we report improved measurements of the color-suppressed decay branching fractions $\mathcal{B}(\bar{B}^{0} \to D^{0}\eta)$ = (26.6 $\pm$ 1.2 $\pm$ 2.1) $\times$ $10^{-5}$ and $\mathcal{B}(\bar{B}^{0} \to D^{0}\bar{K}^{0})$ = (5.6 $\pm$ 0.5 $\pm$ 0.2)  $\times$ $10^{-5}$. The first and second quoted uncertainties are statistical and systematic, respectively.
\end{abstract}

\maketitle
\begin{figure} [!b]
\vspace{-0.8 cm}
  \mbox{
     \subfigure[]{\includegraphics[width=0.40\columnwidth]{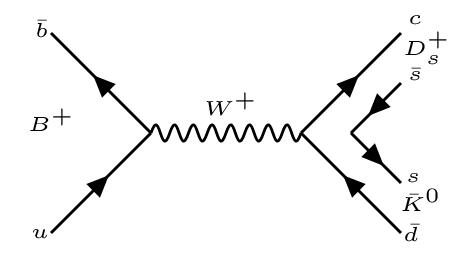}}
     \subfigure[]{\includegraphics[width=0.45\columnwidth]{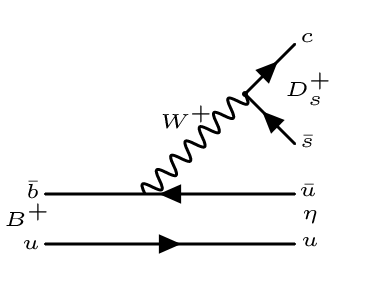}}
  }
  \mbox{
    \subfigure[]{\includegraphics[width=0.45\columnwidth]{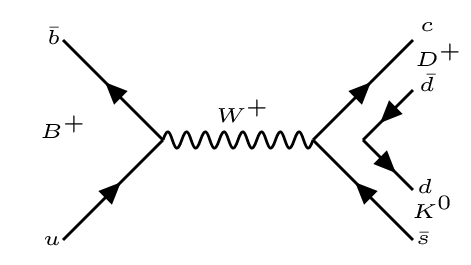}}
    \subfigure[]{\includegraphics[width=0.45\columnwidth]{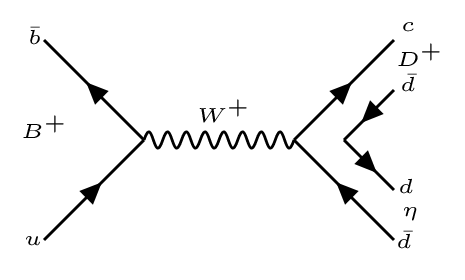}}
    }
\caption{Feynman diagrams for (a) $B^{+} \to D_{s}^{+}\bar{K}^{0}$ annihilation amplitude, (b) $B^{+} \to D_{s}^{+}\eta$ spectator amplitude, (c) $B^{+} \to D^{+}K^{0}$ annihilation amplitude, and (d) $B^{+} \to D^{+}\eta$  annihilation amplitude.}
  \label{fig_btod}
\end{figure}

%\section{\label{sec:level1}First-level heading}
% sections are not used for PRL papers
%Significant $CP$ violation is described by a single irreducible complex phase of the Cabibbo-Kobayashi-Maskawa (CKM) unitarity matrix $V$~\cite{sm} in the Standard Model (SM). It is considered to be too small to produce the observed
%matter-antimatter asymmetry in the universe. Hence, New Physics contributions are required to be search for by testing unitarity conditions for $V$ in a variety of processes. The study of rare $B$ decays could provide hints of the physics beyond the SM~\cite{rare_new}.\\
The dominant amplitude for the decay $B^+ \to D_s^{+} \bar{K}^0$ is expected to be the weak-annihilation process, where the initial-state $\bar{b}u$ pair annihilates to produce a virtual $W^+$ boson as shown in Fig.~\ref{fig_btod} (a). Such annihilation amplitudes cannot be evaluated using the factorization approach~\cite{ckm}. The weak-annihilation amplitude is expected to be proportional to $f_{B}/m_{B}$, where $m_{B}$ and $f_{B}$ are the mass and decay constant of $B$ meson, respectively.
Numerically, $f_{B}/m_{B}$  $\approx$ $\lambda^{2}$~\cite{ckm,lam}, where $\lambda \equiv \sin\theta_c \approx 0.22$~\cite{min_sugg} with $\theta_c$ being the Cabibbo angle. These processes are additionally suppressed by the CKM factor $|V_{ub}|\sim \lambda^3$~\cite{sm} and so the resulting amplitudes are naively of the order of $\lambda^{5}$. Therefore, in most theoretical calculations such amplitudes are neglected. However, rescattering effects from other decay modes might increase the branching fractions of decays dominated by the weak annihilation~\cite{lam}. \par
The related decay $B^{+} \to D_{s}^{+}\eta$, the leading process involves a $b \to u$ quark-level transition as shown in Fig.~\ref{fig_btod} (b), which is suppressed by a factor $\vert V_{ub} \vert$. Searching for these decay modes is crucial in order to improve the theoretical understanding, as they provide an insight into the internal dynamics of the $B$ mesons~\cite{ckm}. These rare decays are sensitive probes for physics beyond the Standard Model, and are not well measured. Such measurements provide a benchmark to search for new physics contributions in loop-dominated processes that would constrain the unitarity triangle. Further, these modes also represent a significant background source for analyses of other rare modes. The decays $B^{+} \to D^{+}K^{0}$ and $B^{+} \to D^{+}\eta$ are of interest as these modes are also dominated by the weak-annihilation diagram as shown in Fig.~\ref{fig_btod} (c, d).\par
The upper limits on the branching fractions of $B^{+} \to D_{s}^{(\ast)+}\eta $ and $ B^{+} \to D_{s}^{(\ast)+}\bar{K^{0}} $ decays were set at the 90$\%$ confidence level by the CLEO collaboration using a sample of $1.16 \times  10^{6}$ $ B\bar{B}$ events~\cite{one}. In addition, the {\sc BaBar} collaboration reported an upper limit on $B^{+} \to D^{+}K^{0}$ decays based on a sample of $226 \times  10^{6}$ $ B\bar{B}$~\cite{ba}. The $B^{+} \to D^{+}\eta$ decay mode has never been searched for. To validate the rare decay modes, we use $\bar{B}^{0} \to D^{0}\eta$ as a control mode for $B^{+} \to D_{s}^{+}\eta$, $D_{s}^{\ast+}\eta$, $D^{+}\eta$  decay modes, and  $\bar{B}^{0} \to D^{0}\bar{K}^{0}$ as a control mode for study the $B^{+} \to D_{s}^{+}\bar{K}^{0}$, $ D_{s}^{\ast+}\bar{K}^{0}$, and $D^{+}K^{0}$ decay modes. These control modes were earlier studied by Belle~\cite{co_be,co_be1} and {\sc BaBar}~\cite{baeta,baKs} using samples containing between 85$\times$10$^{6}$ and 454$\times$10$^{6}$   $B\bar{B}$ events.\par
In this paper, we present studies of the branching fraction of rare decay modes $B^{+} \to D_{s}^{+}\eta$, $B^{+} \to D_{s}^{\ast+}\eta$, $B^{+} \to D_{s}^{+}\bar{K}^{0}$, $B^{+} \to D_{s}^{\ast+}\bar{K}^{0}$, $B^{+} \to D^{+}\eta$, and $B^{+} \to D^{+}K^{0}$; where $D_{s}^{\ast+} \to D_{s}^{+}\gamma$; $D_{s}^{+} \to \phi\pi^{+},~\bar{K}^{\ast0}K^{+},~K_{S}^{0}K^{+}$; $D^{+} \to K^{-}\pi^{+}\pi^{+},~K_{S}^{0}\pi^{+}$; and $\eta \to \gamma\gamma,~\pi^{-}\pi^{+}\pi^{0}$. We also report improved measurements of the branching fractions of color-suppressed decay modes $\bar{B}^{0} \to D^{0}\eta$ and $\bar{B}^{0} \to D^{0}\bar{K}^{0}$; where $D^{0}\to K^{-}\pi^{+}$, $K^{-}\pi^{+}\pi^{+}\pi^{-}$, $K_{S}^{0}\pi^{-}\pi^{+}$, and $K^{-}\pi^{+}\pi^{0}$. Charge conjugate decay modes are included throughout the paper unless explicitly stated otherwise. The results are based on the full sample of $772 \times 10^{6}$ $B$ meson pairs collected by the Belle detector at the KEKB asymmetric-energy $e^{+}e^{-}$ collider~\cite{KEKB}. The first sample of $152 \times 10^{6}$ $B\bar{B}$ events was collected with a 2.0~cm radius beam pipe and a three-layer silicon detector, while the remaining $620 \times 10^{6}$ $B\bar{B}$ pairs were collected with a 1.5~cm radius beam pipe, a four-layer silicon detector and modified drift chamber~\cite{belle_mod2}.\par
The Belle detector is a large-solid-angle spectrometer, which includes a silicon vertex detector (SVD), a 50-layer central drift chamber (CDC), an array of aerogel threshold Cherenkov counters (ACC), time-of-flight scintillation counters (TOF), and an electromagnetic calorimeter (ECL) comprised of 8736 CsI(Tl) crystals located inside a superconducting solenoid coil that provides a 1.5~T magnetic field. An iron flux-return yoke located outside the coil is instrumented to detect $K_{L}^{0}$ mesons and muons. Further details of the Belle experiment can be found elsewhere~\cite{Belle_det}.\par
To validate the analysis procedure, determine efficiencies, and study backgrounds, we use samples of simulated data generated with EvtGen~\cite{evtgen} with QED final-state radiation generated by PHOTOS~\cite{photos}. The detector response is incorporated using GEANT3~\cite{geant}. For background studies, we use five separate simulation samples that include $e^{+}e^{-}\to B\bar{B}$ and $q\bar{q}$ $(q$ = $u$, $d$, $s$, $c)$ events. Each such sample has the same size as the data sample. We perform the analysis using the B2BII software package~\cite{b2bii}, which converts Belle data into a format compatible with the Belle II software framework~\cite{basf2_lib}.\par %Signal events are generated using BASF (Belle Analysis Software Framework)~\cite{two}, while the reconstruction is done using BASF2 (Belle II Analysis Software Framework)~\cite{basf2_ref} .\\
We select tracks consistent with originating from the interaction point by requiring  $d_{r}$ $<$ 0.5~cm and $\vert d_{z} \vert$ $<$ 4.0~cm, where $d_{r}$ and $d_z$ are the track impact parameters in the plane transverse and parallel to the beam axis, respectively. Particle identification of $K^{\pm}$ ($\pi^{\pm}$) candidates is accomplished by combining the information from various subdetectors: ionization energy loss from the CDC, the number of photoelectrons from the ACC, and timing measurements from the TOF. We require the likelihood ratio $\mathcal{L}(K/\pi)$ = $\mathcal{L}_{K}/(\mathcal{L}_{K} + \mathcal{L}_{\pi})$ is required to be greater than 0.2 (less than 0.8) for $K^{\pm}$  ($\pi^{\pm}$) candidates, where $\mathcal{L}(h)$ is the likelihood of a track being consistent with the particle $h$. The efficiency for kaon (pion) identification ranges between 85 to 90$\%$ (87 to 91$\%$) depending on the track momentum with the misidentification rate of a pion (kaon) as a kaon (pion) of  about 11 to 16$\%$ (13 to 16$\%$).\par
For photons, we use the ECL clusters that have energy greater than 50, 100, and 150 MeV in the barrel (32.2$^{\circ}$ $<$ $\theta$ $<$ 128.7$^{\circ}$), forward (12.4$ ^{\circ}$ $<$ $\theta$ $<$ 31.4$^{\circ}$), and backward (130.7$^{\circ}$ $<$ $\theta$ $<$ 155.1$^{\circ}$) regions of the ECL, respectively. To suppress misreconstructed $\eta \to \gamma\gamma$ candidates, we select photon candidates whose energies in the center-of-mass (c.m.) frame are greater than 300~MeV. After implementing the energy requirement on photons, the signal loss fraction is 19.6$\%$, while the background rejection fraction is 56.3$\%$. For photon candidates coming from $D_{s}^{\ast+}$, we  select only those whose energies in the c.m. frame are greater than 110~MeV.\par
We require $\pi^{0}\to\gamma\gamma~(\eta\to\gamma\gamma)$ candidates to have an invariant mass $M_{\gamma\gamma}$ within the range [0.115, 0.155]~GeV$/c ^{2}$ ([0.50, 0.58] ~GeV$/c^{2}$), which corresponds to $\pm$3$\sigma$ about the nominal mass of the $\pi^{0}$ $(\eta$) meson~\cite{PDG}, with $\sigma$ being the mass resolution. We also reconstruct $\eta\to\pi^{+}\pi^{-}\pi^0$ candidates, which are required to have an invariant mass in the range [0.535, 0.560]~GeV$/c ^{2}$. The mass interval corresponds to $\pm 3 \sigma$ about the known $\eta$ mass \cite{PDG}. For the selected $\eta$ and $\pi^{0}$ candidates a mass-constrained fit is performed to improve the momentum resolution. \par
We reconstruct $K_{S}^{0}$ candidates by combining pairs of oppositely  charged particles compatible with originating from a common vertex; both charged particles are assumed to be pions. Further, a multivariate algorithm is used to improve the purity of the sample~\cite{Ks}. The $K_{S}^{0}$ candidates are required to have an invariant mass in the range [0.487, 0.508]~GeV$/c ^{2}$, which corresponds to $\pm$3$\sigma$ about the nominal mass of the $K_{S}^{0}$ meson~\cite{PDG}. We retain only the $\phi$ ($\bar{K}^{\ast0}$) candidates having invariant masses within  14 MeV$/c^{2}$ (100 MeV$/c^{2}$) of their known values~\cite{PDG}.\par
The invariant masses of the $D_{s}^{+}$ candidates are required to be within 13, 15, and 17 MeV$/c^{2}$ of the $D_{s}^{+}$ nominal mass~\cite{PDG} for $\phi\pi^{+}$, $\bar{K}^{\ast0}K^{+}$, and $K_{S}^{0}K^{+}$ decay modes, respectively. The invariant mass of $D^{+}$ candidates is required to be within 15 MeV$/c^{2}$ of the nominal mass of the $D^{+}$~\cite{PDG} mesons for both $K^{-}\pi^{+}\pi^{+}$ and $K_{S}^{0}\pi^{+}$ decay modes. These selection criteria correspond to approximately a $\pm 3\sigma$ window. The $D_{s}^{\ast+}$ candidates are selected from combinations of the $D_{s}^{+}$ and a photon. We require $D^{*+}_s$ candidates to have $\Delta M$  between [0.13, 0.16]~GeV$/c ^{2}$, where $\Delta M$ is the difference between the reconstructed mass of $ D_{s}^{\ast+} $ and $D_{s}^{+}$. The invariant mass of $D^{0}$ meson candidates are required to be within 20, 15, 20, and 35 MeV$/c^{2}$ of the $D^{0}$~\cite{PDG} nominal mass for $K^{-}\pi^{+}$, $K^{-}\pi^{+}\pi^{+}\pi^{-}$, $K_{S}^{0}\pi^{-}\pi^{+}$, and $K^{-}\pi^{+}\pi^{0}$ decay modes, respectively. These selection requirements correspond to a $\pm 3\sigma$ window in mass resolution. To reduce the combinatorial backgrounds that include a poorly reconstructed $\pi^{0}$ candidate in the $D^{0} \to K^{-}\pi^{+}\pi^{0}$ decay mode, we use $\pi^{0}$ candidates with c.m. frame momenta greater than 0.4~GeV$/c$ and invariant masses in the range [0.120, 0.145]~GeV$/c ^{2} $.\par
The $B^{+}$ and $\bar{B}^{0}$ meson decays are reconstructed from $D_{s}^{+}$, $D_{s}^{\ast+}$, $D^{+}$, and $D^{0}$ mesons that are combined with an $\eta$ or a $K_{S}^{0}$ candidate. For the reconstruction of $B$ candidates we utilize two kinematic variables: the energy difference $ \Delta E = E_{B} - E_{\rm beam} $, where $E_{\rm {beam}} $ is the beam energy and $ E_{B} $ is the $B$-candidate energy, both calculated in the c.m. frame; and the beam-constrained mass $M_{\rm bc} = \sqrt{\left({E_{\rm beam}}/{c^{2}}\right)^{2} - \left({\vec{p}_{B}}/{c}\right)^{2}}$, where $\vec{p}_{B}$ is the momentum of the $B$ meson candidate in the c.m. frame. The resolution of $M_{\rm bc}$ is between 2.6$-$4.3 MeV$/c^{2}$ for all decay modes. The resolution of $\Delta E$ depends upon the number of photons in the final state. Candidates satisfying the $ \vert\Delta E \vert < $ 0.18~GeV  and $M_{\rm bc} >$ 5.27~GeV$/c^{2}$ criteria are retained for further consideration. The $\Delta E$ interval is kept wide for two reasons: to take care of the asymmetric signal shape in modes containing an $\eta$, and to model peaking backgrounds effectively. Vertex- and mass-constrained fits are performed on intermediate candidates, such as $D_{s}^{+}$, $D^{0}$, $\phi$, and $K_{S}^{0}$, while only vertex-constrained fits are performed on $B^{+}$, $\bar{B}^{0}$, $\bar{K}^{\ast0}$ candidates. These kinematic fits result in an improved determination of the energy and momenta of the candidate $B$ mesons.\par
 The production cross-section of $e^{+}e^{-} \to q\bar{q}$ is approximately three times that of $B\bar{B}$ production at energies close to the $\Upsilon (4S)$ resonance, making the continuum background suppression necessary in all modes of interest. In the c.m. frame, continuum events generally have particles collimated into back-to-back jets, whereas the particles from the nearly-at-rest $B$ mesons produced in $B\bar{B}$ events are isotropically distributed over the full solid angle. Therefore, we combine event-shape variables and flavor-tagging information using a multivariate classifier FastBDT~\cite{FastBDT_belle} to distinguish between continuum and $B\bar{B}$ events. The FastBDT algorithm uses the following seven variables: two modified Fox-Wolfram moments~\cite{KSFW}; the absolute value of the cosine of the angle between the thrust axis of the $B$ candidate and that of the rest of the event in the c.m. frame; the thrust value of the signal $B$ candidate particles; the CLEO cone~\cite{cleo_cone} in 10$ ^{\circ} $ of the thrust axis of the $B$ candidate; the absolute value of the cosine of the angle between the $B$ candidate momentum and the beam axis in the c.m. frame; and the $B$ meson category-based flavor-tagger output~\cite{fl}. The Belle II flavor taggers are multivariate algorithms that receive track-hit and charged-particle identification information about particles as kinematic input on the tag side, and provide the flavor of the tag-side $B$ meson.
These flavor tagger variables provide additional discrimination in our study to separate  $B\bar{B}$ like events from $q\bar{q}$ events.\par 
The continuum background peak at zero and signal at one in the distribution of FastBDT classifier output ($C$). We do not find any correlation between $C$ and $\Delta E$. We require candidates to have $C>0.92$; this criterion is optimized by maximizing the figure-of-merit defined in Ref.~\cite{three} and retains 52, 40, 55, and 47$\%$ of signal events, while removing approximately 98$\%$ of background events, for $B^{+} \to D_{s}^{+}\eta$, $B^{+} \to D_{s}^{+}K_{S}^{0}$, $B^{+} \to D^{+}\eta$, and $B^{+} \to D^{+}K_{S}^{0}$ decays, respectively. We use the BDT classifiers trained on the signal modes for our study of control modes to validate and calibrate the selection.\par
After the reconstruction, 0.9 $-$ 11$\%$ of events contain multiple $B$ candidates, depending on the decay modes. When there are more than one $B$ candidates in a given event, we select the best candidate (BCS) with the smallest value of $\chi^{2}_{\rm BCS}$, defined as:
\begin{equation}
\chi^{2}_{\rm BCS} = \chi^{2}_{M_{\rm bc}} + \chi^{2}_{M_{i}}, 
\end{equation}
 where the $\chi^{2}_{M_{i}}$ variable is calculated using the reconstructed mass $M_{i}$, its resolution $\sigma_{i}$, and the corresponding nominal mass $m_{i}$~\citep{PDG} of the reconstructed meson $i$ as $\chi^{2}_{M_{i}}$ = $\left( \frac{M_{i}-m_{i}}{\sigma_{i}}\right)^{2} $ and $i$ indicates a $D_{s}^{+}$, $D^{+}$, and $D^{0}$ meson. Table~\ref{tab_res} summarizes the resolution ($\sigma_{i}$) of $D_{(s)}^{+}$, $D^{0}$, and $M_{\rm bc}$ used to estimate $\chi^{2}_{M_{i}}$. The BCS chooses the correctly reconstructed $B$ candidate between 57 $-$ 70$\%$ of the time, depending on the decay mode. \par
\begin{table} [htb]
\caption{The mass resolution ($\sigma_{i}$) of $D_{s}^{+}$, $D^{+}$, $D^{0}$, and $M_{\rm bc}$ used to estimate the $\chi^{2}$ variable.}
\begin{center}
\begin{tabular}{lccc}
\hline
\hline
\multicolumn{4}{c}{Mass resolution (MeV$/c^{2}$) for reconstructed decays}\\
\hline
\multirow{2}{2em}{$\sigma_{D_{s}^{+}}$} &  $D_{s}^{+} \to \phi\pi^{+}$ &  $D_{s}^{+} \to \bar{K}^{\ast0}K^{+}$ &  $D_{s}^{+} \to K_{S}^{0}K^{+}$\\ 
 & 3.8 & 4.0 & 5.1 \\
\hline
\multirow{2}{2em}{$\sigma_{D^{+}}$} & \multicolumn{2}{c}{$D^{+} \to K^{-}\pi^{+}\pi^{+}$}  & $D^{+} \to K_{S}^{0}\pi^{+}$ \\ 
 & \multicolumn{2}{c}{4.7} & 5.1 \\
\hline
\multirow{3}{2em}{$\sigma_{D^{0}}$} & $D^{0} \to K^{-}\pi^{+}$ & \multirow{2}{*}{$D^{0} \to K^{-}\pi^{+}\pi^{+}\pi^{-}$}   & \multirow{2}{*}{$D^{0} \to K^{-}\pi^{+}\pi^{0}$}\\
& $D^{0} \to K_{S}^{0}\pi^{+}\pi^{-}$ &   &  \\
& 6 & 5 & 12 \\
\hline
\multirow{3}{2em}{$\sigma_{M_{\rm bc}}$} & $B^{+} \to D_{s}^{+}\eta$ & $B^{+} \to D_{s}^{+}K_{S}^{0}$ & $B^{+} \to D_{s}^{\ast+}\eta$  \\
&$B^{+} \to D^{+}\eta$ &$B^{+} \to D^{+}K_{S}^{0}$ & $ B^{+} \to D_{s}^{\ast+}K_{S}^{0}$  \\
 & 2.9 & 2.6 & 4.3 \\
\hline
\hline
\end{tabular}
\end{center}
\label{tab_res}
\end{table}
In $B^{+} \to D^{+}\eta$ ($B^{+} \to D^{+}K_{S}^{0}$), the peaking background at $\Delta E \sim -0.16~\mathrm{GeV}$ comprises candidates reconstructed from $\bar{B}^{0} \to D^{+}\rho^{-}$ ($\bar{B}^{0} \to D^{+}K^{\ast-}$) decay modes. For the control mode, the significant	cross-feed contributions come from $\bar{B}^{0} \to D^{\ast0}h$ to the $\bar{B}^{0} \to D^{0}h$ decay mode at $\Delta E \sim -0.16~\mathrm{GeV}$ since the additional photon is not reconstructed. Another peaking background at around $\Delta E \sim -0.16~\mathrm{GeV}$ in the distributions of $\bar{B}^{0} \to D^{0}\eta$ and $\bar{B}^{0} \to D^{0}K_{S}^{0}$ decays arises from charged $B$ meson decays into three final-state particles and $B^{+} \to D^{0}K^{\ast+}$ decay modes, respectively.\par
\begin{figure*}
\includegraphics[width=5.5cm, height = 5.2cm]{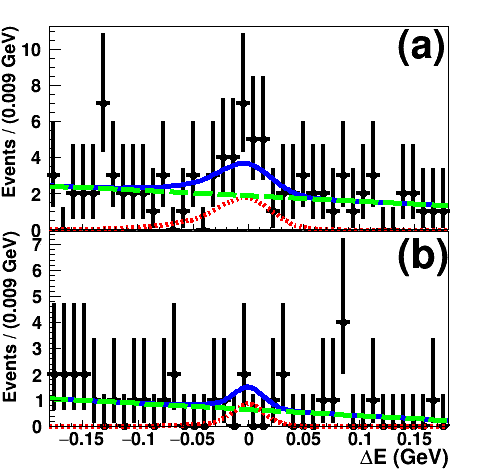}
\includegraphics[width=5.5cm, height = 5.2cm]{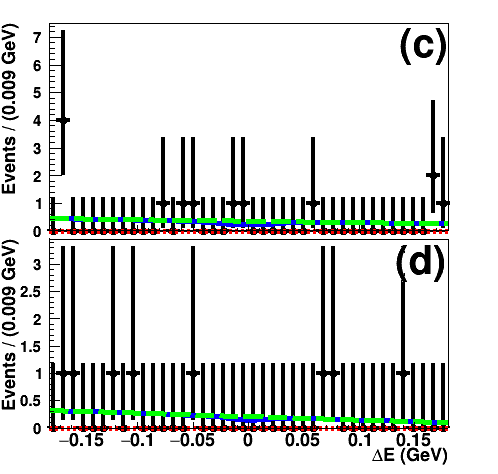}
\includegraphics[width=5.5cm, height = 5.2cm]{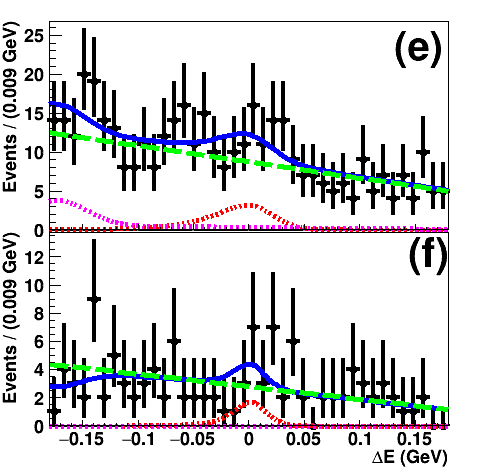}\\
\includegraphics[width=5.5cm, height = 5.2cm]{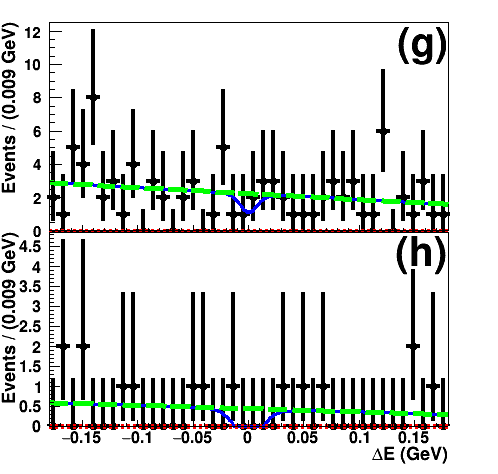}
\includegraphics[width=5.5cm, height = 5.2cm]{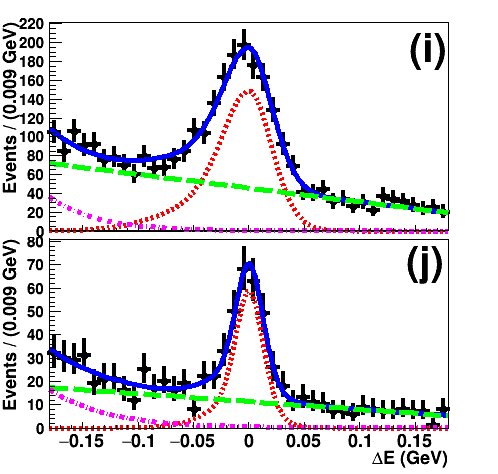}
\includegraphics[width=5.5cm, height = 5.2cm]{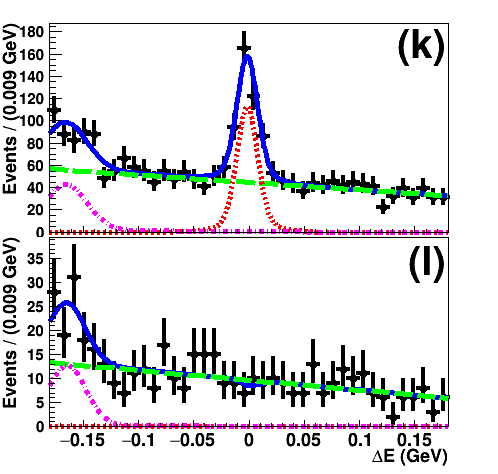}
\caption{\label{fit_dis}Fits to the $\Delta E$ distributions in data for decay modes (a) $B^{+} \to D_{s}^{+}\eta(\gamma\gamma)$, (b) $B^{+} \to D_{s}^{+}\eta(\pi^{-}\pi^{+}\pi^{0})$, (c) $B^{+} \to D_{s}^{\ast+}\eta(\gamma\gamma)$, (d) $B^{+} \to D_{s}^{\ast+}\eta(\pi^{-}\pi^{+}\pi^{0})$, (e) $B^{+} \to D^{+}\eta(\gamma\gamma)$, (f) $B^{+} \to D^{+}\eta(\pi^{-}\pi^{+}\pi^{0})$, (g) $B^{+} \to D_{s}^{+}K_{S}^{0}$, (h) $B^{+} \to D_{s}^{\ast+}K_{S}^{0}$, (i) $\bar{B}^{0} \to D^{0}\eta(\gamma\gamma)$, (j) $\bar{B}^{0} \to D^{0}\eta(\pi^{-}\pi^{+}\pi^{0})$, (k) $\bar{B}^{0} \to D^{0}K_{S}^{0}$, and (l) $B^{+} \to D^{+}K_{S}^{0}$. A simultaneous fit is performed for $\eta \to \gamma\gamma$ and $\eta \to \pi^{-}\pi^{+}\pi^{0}$  decay modes. The black points with error bars show the data points. The different curves correspond to the various fit components: the solid blue curve is the total PDF, the dotted red is the signal PDF, the dash-dotted magenta line is the peaking background PDF, and the green dashed line is the combinatorial background PDF.}  
\end{figure*}

All aforementioned $D$ sub-decay modes are used to reconstruct $B$ candidates except for $\bar{B}^{0} \to D^{0}\eta$, where we exclude the $D^{0} \to K^{-}\pi^{+}\pi^{0}$ sub-decay mode because of the large combinatorial background. The branching fractions of decay modes are extracted from the unbinned maximum-likelihood fits to the $\Delta E$ distributions. For all decay modes, the $\Delta E$ fit is performed in the range $ \vert\Delta E \vert < $ 0.18~GeV. %In all decay modes, the shape of signal, and background events in {blue}{the} $\Delta E$ distribution is fixed from {blue}{fits} to \{blue}{simulation} samples after applying the data/MC correction factor from the control modes. 
For $B^{+} \to D_{s}^{+}\eta$, $B^{+} \to D^{+}\eta$, and $\bar{B}^{0} \to D^{0}\eta$ decay modes, the signal PDF shape in the $\Delta E$ distribution is parametrised with the sum of a Gaussian and a bifurcated Gaussian function with a common mean. For $B^{+} \to D_{s}^{+}K_{S}^{0}$, $B^{+} \to D^{+}K_{S}^{0}$, and $\bar{B}^{0} \to D^{0}K_{S}^{0}$ decay modes, the signal shape is modeled with the sum of two Gaussians with a common mean. The combinatorial background, mainly from continuum events, is modeled with a straight line. The peaking background at $\Delta E\sim -0.16$~GeV from partially reconstructed $B$ decays is modeled with the sum of two Gaussian functions with a common mean. We fix all the parameters of the signal PDF for $B^+ \to D_{s}^{+}h,~B^+ \to D_{s}^{*+}h,~B^+ \to D^{+}h$, and $\bar{B}^{0} \to D^{0}h$  decay modes from the corresponding simulated signal sample after applying a correction for differences between data and simulation in the mean and resolution; the corrections are estimated from the respective control modes. The peaking-background PDF parameters are also fixed to those fit to the generic simulated sample corrected for any resolution and bias with respect to the data as estimated from the control sample. A simultaneous fit is performed for $\eta \to \gamma\gamma$ and $\eta \to \pi^{-}\pi^{+}\pi^{0}$ decay modes for $\eta$ modes in order to account for resolution differences. The projections of fits to the $\Delta E$ distribution are shown in Fig.~\ref{fit_dis}.\par
We calculate the branching fraction using
\begin{equation}
\mathcal{B} = \frac{N_{s}}{N_{B\bar{B}} \times \mathcal{B}(\eta/K_{S}^{0}) \times {\sum_{i}[\varepsilon_{\rm corr_{i}} \times \mathcal{B}_{i}]}},
\end{equation}
where $N _{s} $ is the signal yield from combined $D$ meson sub-decay modes, $N_{B\bar{B}}$ is the number of $B\bar{B}$ events from the data sample ($(772 \pm 10) \times 10^{6}$)~\cite{Belle_det}, $\mathcal{B}_{i}$ is the branching fraction of secondary decays reported in Ref.~\cite{PDG}, and $\epsilon_{corr_{i}}$ is the corrected signal efficiency, where $i$ indicates the different $D$ sub-decay modes. Equation~(2) assumes an equal production of neutral and charged $B$ mesons from $\Upsilon(4S)$. Table~\ref{cor_eff} summarizes the corrected efficiency of the signal modes, as well as the control decay modes.
\begin{table} [htb]
\caption{Summary of the corrected efficiency ($\%$) for the signal and the control decay modes.}
\begin{center}
\begin{tabular}{lccc}
\hline
\hline
Mode & \hspace{-1.2 cm}$D_{s}^{+} (\phi\pi^{+})$ &\hspace{-0.8 cm}$D_{s}^{+} (\bar{K}^{\ast0}K^{+})$ & $D_{s}^{+} (K_{S}^{0}K^{+})$ \\ 
\hline
$B^{+} \to D_{s}^{+}\eta(\gamma\gamma)$ & \hspace{-1.2 cm}5.9 & \hspace{-0.8 cm}6.4 & 6.7 \\
$B^{+} \to D_{s}^{+}\eta (\pi^{-}\pi^{+}\pi^{0})$ & \hspace{-1.2 cm}3.1 & \hspace{-0.8 cm}3.1 & 3.8 \\
$B^{+} \to D_{s}^{\ast+}\eta (\gamma\gamma)$ & \hspace{-1.2 cm}1.8 & \hspace{-0.8 cm}1.4 & 1.2 \\
$B^{+} \to D_{s}^{\ast+}\eta (\pi^{-}\pi^{+}\pi^{0})$ &\hspace{-1.2 cm}0.9 & \hspace{-0.8 cm}0.7 & 0.7 \\
$B^{+} \to D_{s}^{+}K_{S}^{0}$ &  \hspace{-1.2 cm}7.7 & \hspace{-0.8 cm}9.4 & 9.7 \\ 
$B^{+} \to D_{s}^{\ast+}K_{S}^{0}$ & \hspace{-1.2 cm}2.3 & \hspace{-0.8 cm}2.0 & 1.6 \\
\hline
 &   $D^{+} (K^{-}\pi^{+}\pi^{+})$ & & $D^{+} (K_{S}^{0}\pi^{+})$ \\
 \hline
$B^{+} \to D^{+}\eta(\to \gamma\gamma)$ & \hspace{-1.2 cm}8.1 & \hspace{-0.8 cm}& 7.8 \\
$B^{+} \to D^{+}\eta(\to \pi^{-}\pi^{+}\pi^{0})$ & \hspace{-1.2 cm}4.1 & \hspace{-0.8 cm}& 4.5 \\
$B^{+} \to D^{+}K_{S}^{0}$ & \hspace{-1.2 cm}12.3 &\hspace{-0.8 cm} & 13.0 \\
\hline
\end{tabular}
\begin{tabular}{lcccc}
  & $K^{-}\pi^{+}$ & $K^{-}\pi^{+}\pi^{+}\pi^{-}$ & $K_{S}^{0}\pi^{+}\pi^{-}$ & $K^{-}\pi^{+}\pi^{0}$\\
  \hline
$\bar{B}^{0} \to D^{0}\eta(\gamma\gamma)$ & 10.0 & 5.6 & 5.8 & 3.0 \\
$\bar{B}^{0} \to D^{0}\eta(\pi^{-}\pi^{+}\pi^{0})$ & 5.4 & 3.0 & 3.0 & $--$ \\
$\bar{B}^{0} \to D^{0}K_{S}^{0}$ & 15.5 & 8.7 & 8.4 & 4.6 \\
\hline
\hline
\end{tabular}
\end{center}
\label{cor_eff}
\end{table}

\par
Table~\ref{bf_tb} summarizes the yield from the fit, signal significance, and branching fraction obtained from the combined $D_{s}^{+}$, $D^{+}$, and $D^{0}$ sub-decay modes from the fitted distributions of $\Delta E$; the  first and second uncertainties are statistical and systematic, respectively. The signal significance ($\mathcal{S}$) is computed as $\mathcal{S} = \sqrt{2(\textrm{ln}~\mathcal{L}(N_{s}) - \textrm{ln}~\mathcal{L}(N_{s} =0))}$, where $\mathcal{L}(N_{s})$ is the likelihood of the nominal fit and $\mathcal{L}(N_{s} =0)$ is the value obtained after repeating the fit with the signal yield ($N_{s}$) fixed to zero.  % We only get significant yield for $B^{+} \to D_{s}^{+}\eta$ decay modes and
In the absence of a significant yield for signal decay modes,  an upper limit (U.L.) is set on each signal yield at the 90$\%$ confidence level  (C.L.) using a frequentist approach~\cite{fre_app}, which includes systematic uncertainties. We perform pseudo-experiments by generating the fixed background from the final PDF and varying the yield of the input signal. We use the corresponding PDF that has been used to fit data for generating the data sets for pseudo-experiments. The fraction of pseudo-experiments with a fitted yield greater than the estimated signal yield in data has been taken as the confidence level. We also smear the yield in the toys using systematic uncertainties.\par
\begin{table} [htb]
\caption{Summary of the fitted results. Signal yield from the $\Delta E$ fit, significance ($\mathcal{S}$) with systematic included, and measured $\mathcal{B}$; U.L. at 90$ \% $ C.L., where no significant signal is observed. The first (second) uncertainties are statistical (systematic).}
\begin{center}
\begin{tabular}{lccc}
\hline
\hline
Decay Mode & Yield (U.L.) & $\mathcal{S}$ & $\mathcal{B} \times 10^{-5}$ \\ 
\hline
$B^{+} \to D_{s}^{+}\eta$  & 18.4 $\pm$ 7.7 (21) & 1.2 & $<$ 1.4   \\
$B^{+} \to D_{s}^{\ast+}\eta$  &$-$1.45 $\pm$ 2.3 (5.5) & -- & $<$ 1.7  \\
$B^{+} \to D^{+}\eta$  & 34 $\pm$ 16 (41) & 1.4 & $<$ 1.2  \\
$B^{+} \to D_{s}^{+}\bar{K}^{0}$  &$-$2.71 $\pm$ 2.8 (4) & -- & $<$ 0.3    \\
$B^{+} \to D_{s}^{\ast+}\bar{K}^{0}$  &$-$2.64 $\pm$ 1.6 (1.8)  & -- & $<$ 0.6   \\
$B^{+} \to D^{+}K^{0}$  &$-$2.99 $\pm$ 5.7 (8) & -- & $<$ 0.2   \\
$\bar{B}^{0} \to D^{0}\eta$  &  1373 $\pm$ 63  & 24.7 & 26.6 $\pm$ 1.2 $\pm$ 2.1   \\
$\bar{B}^{0} \to D^{0}\bar{K}^{0}$  & 323 $\pm$ 27  & 14.9 & 5.6 $\pm$ 0.5 $\pm$ 0.2  \\
\hline
\hline
\end{tabular}
\end{center}
\label{bf_tb}
\end{table}
Table~\ref{sys_err} summarizes the systematic uncertainties due to  various sources. The dominant source in signal decay modes is the uncertainty on the current world-average values of the secondary decay ($D_{s}^{+}$, $D^{+}$, $\phi$, $\bar{K}^{\ast0}$, $K_{S}^{0}$, $\eta$) branching fractions~\cite{PDG}. The uncertainties related to the PDF shapes are obtained by varying all fixed parameters by $\pm 1\sigma$ and taking the change in the yield  as  the  systematic  uncertainty.
The systematics uncertainty from kaon (pion) identification is estimated from a dedicated $D^{\ast+} \to D^{0}(K^{-}\pi^{+})\pi^{+}$ sample, which is used to correct for the small difference in the signal detection efficiency between simulation and data for the signal decay modes. The uncertainty from $N_{B\bar{B}}$ is 1.4$\%$. The uncertainty on the track finding efficiency is found to be 0.35$\%$ per track. The uncertainties in reconstruction efficiencies of photon and $\eta~(\pi^{0})$ are 3.0$\%$~\cite{eta_corr} and 4.1$\%$~\cite{pi0_error} per particle, respectively. The uncertainty from $K_{S}^{0}$ reconstruction is between 0.1--1.6$\%$,  which is estimated from the calibration factor derived from $D^{\ast \pm} \to D^{0}(K_{S}^{0}\pi^{0})\pi^{\pm}_{\rm slow}$~\cite{ks_error}. The biases of 0.4--23.2$\%$ observed from simplified simulated experiments are also taken as systematics related to the fitting procedure.\par

\begin{table*}
\caption{Systematic uncertainties on pion identification ($\pi$), kaon identification ($K$), tracking,  N$_{B\bar{B}}$, $K_{S}^{0}$ reconstruction, $\eta$ ($\pi^{0}$) reconstruction, photon detection, uncertainty in secondary $\mathcal{B}$, and PDF used for signal extraction.}
\begin{ruledtabular}
\begin{center}
\begin{tabular}{lccccccccccc}
\multicolumn{11}{c}{\textbf{Uncertainty ($\%$)}}\\
\hline
Decay Mode & $\pi$ & $K$ & Tracking & N$_{B\bar{B}}$ & $K_{S}^{0}$ & $\eta$ ($\pi^{0}$) & $\gamma$ & Secondary $\mathcal{B}$ & Signal extraction PDF & Fit bias & Total\\
\hline
$B^{+} \to D_{s}^{+}\eta$ & 0.8 & 1.1 & 1.2 & 1.4 & 0.2 & 4.1 & -- & 2.0 & 0.5 & 8.4 & 9.8\\
$B^{+} \to D_{s}^{\ast+}\eta$ & 1.1 & 1.6 & 1.2 & 1.4 & 0.2 & 4.1 & 3.0 & 2.2 & +4.4, -4.0 & 23.2 & +24.4, -24.3\\
$B^{+} \to D^{+}\eta$ & 1.4 & 0.5 & 1.2 & 1.4 & 0.1 & 4.1 & -- & 1.6 & +9.7, -12.1 & 4.1 &+11.6, -13.7\\
$B^{+} \to D_{s}^{+}K_{S}^{0}$ &  0.3 & 0.7 & 1.8 & 1.4 & 1.6 &-- & --& 1.9 & 1.2 & 2.2 & 4.3\\
$B^{+} \to D_{s}^{\ast+}K_{S}^{0}$ & 0.4 & 0.9 & 1.8 & 1.4 & 1.6 & -- &  3.0 & 2.1 & 1.2 & 9.9 & 11.0\\
$B^{+} \to D^{+}K_{S}^{0}$ & 0.6 & 0.3 & 1.8 & 1.4 & 1.5 &-- &-- & 1.5 & 1.2 & 1.3 & 3.6\\
%\hline
%$\bar{B}^{0} \to D^{0}\eta(\gamma\gamma)$ & 1.1 & 0.7 & 1.0 & 1.4 & 0.1 & 2.1 & 1.2 & 2.0 & +5.8, -5.3 & 0.4  & +7.0, -6.5\\
%\hline
%$\bar{B}^{0} \to D^{0}\eta(\pi^{-}\pi^{+}\pi^{0})$ & 3.0 & 0.6 & 1.9 & 1.4 & 0.2 & 2.1 & 5.7 & 2.0 &  2.8 & 0.6 & 8.0 \\
$\bar{B}^{0} \to D^{0}\eta$& 1.4 & 0.7 & 1.2 & 1.4 & 0.1 & 5.7 & -- & 2.0 & +4.4, -4.0 & 0.6  & +7.9, -7.7\\
$\bar{B}^{0} \to D^{0}K_{S}^{0}$ & 0.8 & 0.5 & 1.7 & 1.4 & 1.5 & 2.0 & -- & 1.9 & 1.2 & 0.4 & 4.1\\
\end{tabular}
\end{center}
\label{sys_err}
\end{ruledtabular}
\end{table*}

In summary, we have searched for $B^{+} \to D_{s}^{+}h^{0}$, $B^{+} \to D_{s}^{\ast+}h^{0}$, and $B^{+} \to D^{+}h^{0}$ decays using the full $\Upsilon(4S)$ data sample recorded by the Belle experiment. In the absence of a significant signal yield, an upper limit at the 90$\%$ confidence level is given for each signal decay mode. We present the first search result for the $B^{+} \to D^{+}\eta$ decay mode. The obtained upper limits are 20 times more stringent than the previous one. We report the most precise measurement to date of the branching fraction for the $\bar{B}^{0} \to D^{0}\bar{K}^{0}$~\cite{baKs,co_be1} decay, which supersedes the previous Belle result~\cite{co_be1}. The branching fraction measurement of $\bar{B}^{0} \to D^{0}\eta$ decay modes is consistent with the world average and supersedes the previous Belle~\cite{co_be} result. \par
This work, based on data collected using the Belle detector, which was
operated until June 2010, was supported by 
the Ministry of Education, Culture, Sports, Science, and
Technology (MEXT) of Japan, the Japan Society for the 
Promotion of Science (JSPS), and the Tau-Lepton Physics 
Research Center of Nagoya University; 
the Australian Research Council including grants
DP180102629, % Sevior
DP170102389, % Varvell
DP170102204, % Yabsley
DE220100462, % Hsu
DP150103061, % Urquijo
FT130100303; % Urquijo;
Austrian Federal Ministry of Education, Science and Research (FWF) and
FWF Austrian Science Fund No.~P~31361-N36;
the National Natural Science Foundation of China under Contracts
No.~11675166,  %Wen-Biao Yan
No.~11705209;  %Yi-Ming Li
No.~11975076;  %Chengping Shen
No.~12135005;  %Chengping Shen 
No.~12175041;  %Xiaolong Wang
No.~12161141008; %Chengping Shen
Key Research Program of Frontier Sciences, Chinese Academy of Sciences (CAS), Grant No.~QYZDJ-SSW-SLH011; % Chang-Zheng Yuan
Project ZR2022JQ02 supported by Shandong Provincial Natural Science Foundation;
the Ministry of Education, Youth and Sports of the Czech
Republic under Contract No.~LTT17020;
the Czech Science Foundation Grant No. 22-18469S;
Horizon 2020 ERC Advanced Grant No.~884719 and ERC Starting Grant No.~947006 ``InterLeptons'' (European Union);
the Carl Zeiss Foundation, the Deutsche Forschungsgemeinschaft, the
Excellence Cluster Universe, and the VolkswagenStiftung;
the Department of Atomic Energy (Project Identification No. RTI 4002) and the Department of Science and Technology of India; 
the Istituto Nazionale di Fisica Nucleare of Italy; 
National Research Foundation (NRF) of Korea Grant
Nos.~2016R1\-D1A1B\-02012900, 2018R1\-A2B\-3003643,
2018R1\-A6A1A\-06024970, RS\-2022\-00197659,
2019R1\-I1A3A\-01058933, 2021R1\-A6A1A\-03043957,
2021R1\-F1A\-1060423, 2021R1\-F1A\-1064008, 2022R1\-A2C\-1003993;
Radiation Science Research Institute, Foreign Large-size Research Facility Application Supporting project, the Global Science Experimental Data Hub Center of the Korea Institute of Science and Technology Information and KREONET/GLORIAD;
the Polish Ministry of Science and Higher Education and 
the National Science Center;
the Ministry of Science and Higher Education of the Russian Federation, Agreement 14.W03.31.0026, % from 15.02.2018
and the HSE University Basic Research Program, Moscow; % from 15.04.2021
University of Tabuk research grants
S-1440-0321, S-0256-1438, and S-0280-1439 (Saudi Arabia);
the Slovenian Research Agency Grant Nos. J1-9124 and P1-0135;
Ikerbasque, Basque Foundation for Science, Spain;
the Swiss National Science Foundation; 
the Ministry of Education and the Ministry of Science and Technology of Taiwan;
and the United States Department of Energy and the National Science Foundation.
These acknowledgements are not to be interpreted as an endorsement of any
statement made by any of our institutes, funding agencies, governments, or
their representatives.
We thank the KEKB group for the excellent operation of the
accelerator; the KEK cryogenics group for the efficient
operation of the solenoid; and the KEK computer group and the Pacific Northwest National
Laboratory (PNNL) Environmental Molecular Sciences Laboratory (EMSL)
computing group for strong computing support; and the National
Institute of Informatics, and Science Information NETwork 6 (SINET6) for
valuable network support.

\end{document}